\documentclass[preprint,njp]{revtex4}
\usepackage{amsmath}
\usepackage{amssymb}
\usepackage{graphicx}

\begin{document}
\title{Electric Aharonov-Bohm effect without a loop in a Cooper pair box}
\author{Young-Wan Kim}
\author{Kicheon Kang}
\email{kicheon.kang@gmail.com}
\affiliation{Department of Physics, Chonnam National University, Gwangju 61186, 
 Republic of Korea}

\begin{abstract}
We predict the force-free scalar Aharonov-Bohm effect of a Cooper pair box 
in an electric field at a distance
without forming a closed path of the interfering charges. 
The superposition of different charge states plays a major role in 
eliminating the closed loop, which is distinct from the original
topological Aharonov-Bohm effect.
The phase shift is determined 
by the charge-state-dependent local field interaction energy. 
In addition, our proposed setup does not require a pulse experiment
for fast switching of a potential,
which eliminates the major experimental obstacle for observing 
the ideal electric Aharonov-Bohm effect.
\end{abstract}

\maketitle

\section{Introduction}
A charge moving in an external electromagnetic field exhibits 
topological quantum interference known as the 
Aharonov-Bohm~(AB) effect~\cite{aharonov59}.
An intriguing aspect of the
AB effect is that the appearance of a phase shift does not require 
local overlap of the particle and external field. 
For this reason, the AB effect has been
regarded as a pure topological phenomenon that cannot be described 
in terms of the local actions of physical variables.
This property also implies that a loop geometry is essential 
for its observation. 
In the case of the electric Aharonov-Bohm (EAB) effect, 
interference can be observed without an overlap of the particle
and the external electric field, as shown in the original
work by Aharonov and Bohm~\cite{aharonov59}. 
The wave packet of a moving charge splits into two parts and
each part propagates under different scalar potentials. The recollected wave
would show interference with its phase shift proportional 
to the potential difference. 
Unlike the magnetic AB effect, 
a fast switching of the scalar potential is necessary to avoid
overlap of the particle and the external field, which is the main
technical obstacle for an observation of the ideal EAB effect.
The predicted EAB effect is also
topological, and is generally believed to indicate the physical significance
of the (scalar) potential.  
The Aharonov-Casher~(AC) effect~\cite{aharonov84}, 
a related topological quantum
phenomenon, describes the phase shift of a neutral particle with 
a magnetic moment moving around a charged rod~(Fig.1~(a)). 
The AC effect can be regarded as the dual of the
magnetic AB effect in that the roles of the charge and magnetic flux 
(moment) are reversed~(see, e.g., 
Refs.~\onlinecite{reznik89} and \onlinecite{kang15}). 
For the AC effect, a closed loop of the particle's path
is not always required for its realization. For instance,  
the AC phase shift can be observed in the interference of 
two opposite magnetic moments of a neutral particle without dividing 
the particle's path~(Fig.~1(b)).
 This ``loop-free" interference 
had been predicted by Anandan~\cite{anandan82} before the prediction of
the topological AC effect
and was experimentally demonstrated~\cite{sangster93}.
Interference can be achieved, as the AC interaction Lagrangian~(and 
the phase accumulation) depends on the magnetic moment.
Interestingly, this type of interference appears in a different context, 
namely, in
electronic devices with Rashba spin-orbit interactions~(see e.g., 
Refs.~\onlinecite{datta90,shekhter16,tomaszewski18,aharony18}).

Together with the loop-free AC interference, 
the duality of the AB and AC effects 
poses an interesting question
as to whether we can find an AB analogue of loop-free interference.
Exchanging the roles of the magnetic moment
and electric charge in the loop-free AC effect, 
the corresponding loop-free AB effect should appear.
Basically, this is possible in a superposed state of different charges, 
as the AB interaction depends on the charge of the interfering particle. 
However, two key issues should be resolved in order to achieve 
loop-free AB interference.
First, an ordinary charged particle cannot form a superposition 
of different number states, and is inappropriate for our purpose.
This problem can be overcome by utilizing a superconducting 
condensate composed of the superposition of different numbers of Cooper pairs. 
The second problem is that loop-free interference 
cannot be described by a potential difference across two
different positions (which is the case in the topological AB effect), 
as the test particle's wave packet would neither split 
nor form a closed path. 
Instead, the phase shift should be determined by the electrostatic energy difference
(in the electric AB effect) 
between different charge states {\em at the same position}.
We will show that this energy difference is described by the {\em geometric
potential}, defined on the basis of the ``Lorentz-covariant field interaction
(LCFI)" 
approach~\cite{kang13,kang15}, whereas the magnitude of the energy
difference is ambiguous in the conventional potential-based framework. 
A single Cooper pair box (SCB)~(see, e.g., Ref.~\onlinecite{makhlin01}), 
composed of a superposed state of two
different charges, is an ideal system for its observation.
We predict a loop-free electric Aharonov-Bohm~(EAB) effect in an SCB; 
a relative phase shift between two charge states appears 
in an SCB influenced by an external electric field at a distance. 
The magnitude of the phase shift is proportional to 
difference in the electrostatic energies between the two charge states.
In addition, we point out that the EAB effect under the ideal condition
- that the charge and external field does not overlap -
can be more easily realized in this setup, 
as it does not require fast switching of a potential,
the major technical difficulty for realizing the original force-free EAB 
interference in a two-path interferometer.

The paper is organized as follows. In Section II, the original EAB
effect is reinvestigated in the framework of the LCFI Lagrangian,
which shows that the local approach reproduces the 
prediction for the original topological EAB effect. 
Section III is devoted to our main prediction 
of the loop-free EAB effect for a Cooper pair box.
Discussion and Conclusion are given in Sections IV and V, respectively.

\section{Field interaction approach to the electric Aharonov-Bohm effect.}
Let us begin by briefly reviewing the original EAB effect in an ideal
situation~(see Fig.~2(a))~\cite{aharonov59}. 
The wave packet of an incident particle with 
charge $q$ 
splits into two parts and enter long Faraday cages. 
In each part, the electrical potential $V_i$ ($i=1,2$) is
switched on after the wave packet enters the cage. 
The duration of the voltage pulse should be sufficiently short to ensure
that each potential is switched off far before the particle exits.
The purpose of this arrangement is to avoid local overlap 
of the particle and electromagnetic field. 
Let the wave function $\psi_0(\mathbf{x},t) = 
\psi_1(\mathbf{x},t)+\psi_2(\mathbf{x},t)$ in the absence of the 
potentials, where $\psi_1$ and $\psi_2$ represent the two parts. 
The electric potential modifies the wave function as
\begin{equation}
 \psi(\mathbf{x},t) = \psi_1(\mathbf{x},t)e^{-iq\int V_1 dt/\hbar}  
                    + \psi_2(\mathbf{x},t)e^{-iq\int V_2 dt/\hbar}  \,,
\end{equation}
and the interference fringe is determined by the phase difference
\begin{equation}
 \varphi = -\frac{q}{\hbar}\int V_0 dt ,
\label{eq:phi_pot}
\end{equation}
where $V_0=V_1-V_2$ is the potential difference.

This EAB effect can also be described in an alternative LCFI 
approach~\cite{kang13,kang15}. 
The essence of this approach is summarized as follows.
The Lagrangian governing the interaction between
a charge and an external field is universally represented by the
local overlap between the external field and that generated by the charge,
instead of the charge being influenced by the external potential. 
Incorporating the Lorentz covariance and linearity in field strengths,
we can uniquely determine the interaction Lagrangian:
\begin{equation}
 L_{in} = \frac{1}{8\pi} \int F_{\mu\nu}^{(q)}F^{\mu\nu} d^3\mathbf{r}'\,,
\end{equation}
where $F^{\mu\nu}$ and $F_{\mu\nu}^{(q)}$ are 
the external electromagnetic field tensor and 
that generated by the charge, respectively.
This Lagrangian reproduces the results derived from the
potential-based approach for the classical equation of motion
and the topological AB effect~\cite{kang13,kang15}. 
In our arrangement of the charge with an external electric field
(in the absence of an external magnetic field), 
the interaction Lagrangian is reduced to 
\begin{equation}
 L_{in} = -U_q =
 -\frac{1}{4\pi} \int \mathbf{E}_q\cdot\mathbf{E}\, d^3\mathbf{r}' \,,
\label{eq:Lin}
\end{equation}
where $U_q$ denotes the energy produced by the interaction between 
two electric fields: the external $\mathbf{E}$
and $\mathbf{E}_q$ produced by the moving charge. 

Fig.~2(b) shows a possible configuration of the external 
electric field when the
potentials $V_1$ and $V_2$ (of Fig.~2(a)) are switched on. 
The essential condition of a nonoverlapping particle and
$\mathbf{E}$ is satisfied. Nevertheless, their interaction 
is manifested in the overlap of $\mathbf{E}$ with
$\mathbf{E}_q$ (not with the position of the charge) in the Lagrangian 
of Eq.~\eqref{eq:Lin}. The moving charge $q$ with speed $u$ 
along the $x$ axis generates the electric field
\begin{equation}
 \mathbf{E}_q(\mathbf{r}) 
  = q\frac{ \gamma[(x-ut)\hat{\mathbf{x}} + y\hat{\mathbf{y}} 
    + z\hat{\mathbf{z}}] }{ [\gamma^2(x-ut)^2+y^2+z^2]^{3/2} } \,,
\label{eq:Eq}
\end{equation}
where $\gamma=1/\sqrt{1-(u/c)^2}$.
For the lower path (region I), we find from
Eqs.~\eqref{eq:Lin} and \eqref{eq:Eq} that 
\begin{equation}
 L_{in} = L_{in}^\mathrm{I} = -\frac{qV_0}{2}\,,
\label{eq:Lin_I}
\end{equation}
where $V_0 = \int \mathbf{E}\cdot d\mathbf{x} = V_1-V_2$ is the potential drop 
across the two regions. Notably, $L_{in}$ is independent of the speed of the
moving charge.
Similarly, for the upper path (region II), we obtain 
\begin{equation}
 L_{in} = L_{in}^\mathrm{II} = \frac{qV_0}{2}\,.
\end{equation}
Therefore, the phase difference accumulated by the interaction is 
\begin{equation}
 \varphi = \frac{1}{\hbar} \int (L_{in}^\mathrm{I}-L_{in}^\mathrm{II})\, dt
   = -\frac{q}{\hbar}\int V_0 dt \,,
\label{eq:phi_lcfi}
\end{equation}
demonstrating that the EAB phase shift (Eq.~\eqref{eq:phi_pot}) is
reproduced in the field interaction approach.
Here, we have considered infinite planar conducting plates generating
the potential difference,
but the phase shift in Eq.~\eqref{eq:phi_lcfi} can be verified
for an arbitrary geometry of $\mathbf{E}$ with the potential difference $V_0$ 
across the two regions.


\section{Electric Aharonov-Bohm effect in a Cooper pair box and the
geometric potentials.}
%
%
Next, we demonstrate how the loop-free EAB effect appears in 
a superposed state of different charges. 
An SCB~\cite{makhlin01},
an artificial two-level quantum system composed of superconducting
circuits, is an ideal system for realization of the loop-free EAB effect.
Its quantum state $|\psi_0(t)\rangle$ is composed of a superposition of
two different charge states $|q\rangle$ and $|q'\rangle$:
\begin{equation}
 |\psi_0(t)\rangle = a(t)|q\rangle + b(t)|q'\rangle \,,
\label{eq:psi_0}
\end{equation}
where $q-q'=2e$, implying that $|q\rangle$ contains an extra Cooper pair
than $|q'\rangle$.
As discussed above, the charge in the SCB interacts with 
an external electric field ($\mathbf{E}$) at a 
distance~(see the various configurations shown in Fig.~3).
The quantum dynamics of an SCB is more complicated than the
case of the EAB effect in a two-path interferometer. Nevertheless, it is
instructive to analyze the limit of negligible charge transfer between $|q\rangle$
and $|q'\rangle$. 
In this limit, 
the quantum state evolution (modified by the distant electric field)
\begin{equation}
 |\psi(t)\rangle \simeq a(t) e^{-iq V t/\hbar} |q\rangle 
                 + b(t) e^{-iq' V t/\hbar} |q'\rangle 
\end{equation}
is equivalent to that of a two-path interferometer.
$V$ is the scalar potential at the position of the SCB.
Switching the voltage is not required here, 
in contrast to the original EAB setup of moving charges.
The relative phase shift is given by
\begin{equation}
 \varphi = -\frac{(q-q')}\hbar  V t 
         = -\frac{2e}\hbar  V t \,.
\label{eq:phi_cpb_pot}
\end{equation}
The problem with this result is that, 
unlike the original EAB phase given in
Eq.~\eqref{eq:phi_pot}, the phase shift of
Eq.~\eqref{eq:phi_cpb_pot} remains undetermined,
as $V$ at a single position is not a quantity with a definite value.

This ambiguity is removed by adopting the LCFI Lagrangian 
of Eq.~\eqref{eq:Lin}. 
Considering that the field interaction depends on the charge state,  
we can obtain
a well-defined phase shift.   
The interaction Lagrangian for charge $q$ and external $\mathbf{E}$ 
of Eq.~\eqref{eq:Lin}
can be rewritten in an instructive form
\begin{subequations}
\begin{equation}
 L_{in} = -U_q = -q V_G \,,
\label{eq:U_q}
\end{equation}
where 
\begin{equation}
 V_G = \frac{1}{4\pi} 
  \int \frac{\mathbf{E}(\mathbf{r})\cdot\hat{\mathbf{r}} }{ r^2 }\, 
       d^3\mathbf{r} 
\label{eq:V_G}
\end{equation}
\label{eq:UqVG}
\end{subequations}
is a type of scalar potential determined
by the overall distribution of $\mathbf{E}$. 
This ``geometric potential" ($V_G$) plays a similar role as an 
electric scalar potential but is different
from the latter, as $V_G$ at a given position
is uniquely determined by the distribution of $\mathbf{E}$. 
Therefore, the charge-state-dependent phase shift is also uniquely
determined. From Eq.~\eqref{eq:UqVG}
we obtain the state evolution
\begin{equation}
 |\psi(t)\rangle \simeq a(t) e^{-iq V_G t/\hbar} |q\rangle
                 + b(t) e^{-iq' V_G t/\hbar} |q'\rangle 
\label{eq:psi}
\end{equation}
and the well-defined relative phase shift
\begin{equation}
 \varphi = -\frac{2e}\hbar  V_G t \,.
\label{eq:phi-eab}
\end{equation}
This constitutes the EAB effect in an SCB without forming
a loop of the particle's path. 
%
%

As the EAB phase shift is entirely determined by $V_G$, it will be useful
to evaluate its values for different cases~(see Fig.~3).
First, consider a capacitor composed of a pair of two infinite parallel 
conducting plates~(Fig.~3(a)). 
The field interaction energy
is equivalent to that obtained in Eq.~\eqref{eq:Lin_I}, and we find that
\begin{equation}
  V_G = \frac{V_0}{2} \,,
\end{equation}
where $V_0$ is the voltage drop across the two plates.
Similarly, in the presence of two such parallel capacitors~(Fig.~3(b)), 
\begin{equation}
 V_G = (V_0+V_0')/2 
\end{equation}
for the two voltage drops $V_0$ and $V_0'$ 
(measured from the SCB position) across the  
upper and lower capacitors, respectively.

In fact, $V_G$ can be evaluated for a capacitor 
with an arbitrary shape~(Fig.~3(c)). 
The external field can be written as
$\mathbf{E} = E_0\hat{\mathbf{n}}$, where $\hat{\mathbf{n}}$
is the unit vector perpendicular to the capacitor surface,
and Eq.~\eqref{eq:V_G} is reduced to
\begin{subequations}
\begin{equation}
 V_G = \frac{1}{4\pi} \int E_0 \hat{\mathbf{n}}\cdot\hat{\mathbf{r}}\, dr 
       d\Omega \,. 
\end{equation}
By noting that the potential drop across the capacitor can be expressed as
\begin{equation}
 V_0 = \int E_0 \hat{\mathbf{n}}\cdot\hat{\mathbf{r}}\, dr \,,
\end{equation}
we obtain the relation
\begin{equation}
 V_G = \frac{\Omega}{4\pi} V_0 \,,
\label{eq:V_G3}
\end{equation}
\end{subequations}
where $\Omega$ is the solid angle formed by the capacitor geometry.
In all cases of Fig.~3, the EAB phase shift in Eq.~\eqref{eq:phi-eab}
is determined 
by the geometry-dependent potential $V_G$ and not by the voltage difference $V_0$.
This result is in contrast with the original EAB effect where only the voltage
difference of two interfering paths matters.

The EAB effect discussed above can be demonstrated 
in a realistic SCB circuit~(see Fig.~4).
An experimental SCB circuit is controlled
by the gate voltage $V_g$. The circuit has two
capacitances, the junction capacitance $C_J$ and gate capacitance $C_g$.
A new component here is the inclusion of the geometric potential ($V_G$) 
associated with the electric field that is spatially separated from the circuit.

The electrostatic energy of this system has the form
\begin{equation}
 \frac{C_J}{2}V_J^2 + \frac{C_g}{2} (V_g-V_J)^2 + qV_G \,,
\end{equation}
where $V_J$ is the voltage across the tunnel junction.  
Josephson coupling leads the relation between the phase variable ($\phi$) and
$V_J$: $\dot{\phi} = 2eV_J/\hbar$. 
The island charge is 
$q = C_JV_J - C_g(V_g-V_J) = C_\Sigma V_J - C_gV_g$, where $C_\Sigma = C_J+C_g$ is the
total capacitance. Including Josephson coupling (with the constant $E_J$)
as well, 
the Lagrangian of the system is given by (omitting constant terms)
\begin{equation}
 L(\phi,\dot{\phi}) = \frac{C_\Sigma}{2} 
   \left[ \frac{\hbar}{2e}\dot{\phi} - \frac{1}{C_\Sigma}(C_gV_g-C_\Sigma V_G)  
   \right]^2
 + E_J\cos{\phi} \,.
\end{equation}
Adopting the standard procedure of the Legendre transformation,
we obtain the Hamiltonian 
\begin{equation}
  H = E_c(\hat{n}-n_G-n_g)^2 - E_J\cos{\phi} \,,
\label{eq:H}
\end{equation} 
where $E_c=(2e)^2/2C_\Sigma$ is the charging energy of a single Cooper pair.
The number of excess Cooper pairs ($\hat{n}$) of the island
satisfies the commutation rule $[\phi,\hat{n}] = i$
and is limited to $\hat{n}=0,1$ in an SCB.
The effects of the geometric and gate voltages are included in the
variables $n_G = C_\Sigma V_G/2e$ and $n_g = -C_gV_g/2e$, respectively. 

The eigenvalues of this Hamiltonian are 
\begin{subequations}
\begin{equation}
 E_\pm = \mp \frac{1}{2} \sqrt{ \left[ E_c(1-2n_g) - 2eV_G \right]^2 + E_J^2 } \,;
\end{equation}
therefore, the evolution of the quantum state depends on $V_G$.
Fig.~5 displays the qubit spectra both 
with and without the geometric potential.
The spectrum is shifted in the presence of the geometric
potential ($V_G$), and this is the result of the interaction of the qubit   
charge
with the electric field at a distance (as described in Eq.~\eqref{eq:UqVG}).
Accordingly, the transition frequency 
\begin{equation}
 \omega = \frac{1}{\hbar} (E_- - E_+)
  = \frac{1}{\hbar}\sqrt{ \left[ E_c(1-2n_g) - 2eV_G \right]^2 + E_J^2 }
\end{equation}
\label{eq:E_pm}
\end{subequations}
also depends on $V_G$. 
The EAB effect manifested in Eq.~\eqref{eq:E_pm} can be
probed in the standard SCB circuit~\cite{makhlin01,wendin07}. 
The only new component here is to incorporate an external electric field
spatially separated from the circuit. The effect will become prominent when
$2eV_G$ becomes comparable to $E_C$. Typical value of charging energy
is $E_C \sim 100 \rm{\mu eV}$~\cite{nakamura99,astafiev04}, 
and therefore
a variation of the voltage drop $0\leq V_0 \lesssim 100 \rm{\mu V}$ 
(e.g., between the conducting plates in Fig.~3(a))
would be ideal for observation
of the EAB effect. As the measurement techniques in 
the superconducting qubits are well established 
both in the energy- and the time-domain experiments 
(see e.g., Refs.~\onlinecite{makhlin01,wendin07}), 
exploring
the EAB effect of Eq.~\eqref{eq:E_pm} is possible with 
the standard measurement scheme.
For example, measurement of the qubit state can be implemented by connecting
the SCB to an electrometer composed of a single electron transistor (SET)
(see Fig.~6). The state readout can be performed by measuring the 
qubit-state-dependent current through the SET~\cite{wendin07}. 

\section{Discussion.}
Let us discuss several notable aspects of our results. 
First, the loop-free EAB
effect in an SCB cannot be properly accounted for by 
the conventional scalar potential, as the latter does not provide a definite value 
at a given position~(see Eq.~\eqref{eq:phi_cpb_pot}).
The effect is instead described by the 
geometric potential $V_G$~(Eq.~\eqref{eq:V_G}), 
which is determined by the geometry of the external field distribution
and not simply by the potential difference between different positions.
In other words, the EAB effect in our arrangement is understood only by specifying 
the local overlap of the external field and the field generated 
by the interfering charge~(Eq.~\eqref{eq:Lin}), 
demonstrating the locality of the interactions.


Second, consider charge redistribution on the
conducting plate induced by the SCB charge $q$~(Fig.~7). 
This may influence
the interaction between $\mathbf{E}_q$ and
$\mathbf{E}$, and its consequences should be clarified.
For simplicity, the conductor is assumed to be ideal; 
the charges on its surface are free to move in response to $q$. 
A naive expectation would be that the field $\mathbf{E}_q$ generated 
by $q$ is compensated by the field $\mathbf{E}_i$ generated
by the induced charges; $\mathbf{E}_i+\mathbf{E}_q=0$. 
If this is the case, the interaction between $\mathbf{E}_q$ and the external field $\mathbf{E}$
would be completely removed. 
This would result in the disappearance
of the EAB effect in the SCB. 

However, this naive expectation is incorrect, as the quantum
nature of $\mathbf{E}_i$ is not taken into account. 
In a quantum mechanical treatment,
only the expectation value of $\mathbf{E}_i+\mathbf{E}_q$ 
vanishes, whereas the interaction between $q$ and the external field is not
shielded at all, as we show below (see also Ref.~\onlinecite{kang15}). 
The charged particles in the conductor contribute to the field interaction
Lagrangian as 
\begin{equation}
 L_{in}' = - \sum_j q_j V_G(\mathbf{r}_j)  \,,
\end{equation} 
where $q_j$ and $V_G(\mathbf{r}_j)$ are the charge and geometric potential 
(defined as Eq.~\eqref{eq:V_G}) at position $\mathbf{r}_j$ on the 
conductor. For a capacitor with an arbitrary shape~(Fig.~3(c)),
the geometric potential~(Eq.~\eqref{eq:V_G3}) depends only on the 
solid angle formed by the capacitor and is independent of $\mathbf{r}_j$. 
Therefore, the interaction Lagrangian $L_{in}' = - (\sum_j q_j) V_G$ 
is independent of the redistribution of the particles in the conductor,
implying that charge redistribution does not affect the interactions 
between $\mathbf{E}_q$ and $\mathbf{E}$ at all.
The EAB interference is unaffected by the 
induced charges of the conductor.
We can equally apply this argument to the original topological EAB effect.
In addition, the interaction between $\mathbf{E}_q$ and $\mathbf{E}_i$
also does not affect the EAB effect, as it is independent of $\mathbf{E}$.
%

%
Third, although our study is focused on the simpler EAB effect,
it is also possible to demonstrate a magnetic AB effect without a loop.
Moving particles with superposed charge states are necessary 
to achieve it.
This can be realized, for example, by utilizing 
the Andreev reflections 
in superconductor-metal hybrid junctions~\cite{kang17}. 

%
Finally, note that the EAB experiment of the original form 
(as in Fig.~2(a))
has never been performed. 
This is primarily because its realization would require
extremely fast switching of the electric potential 
at one of the Faraday cages placed along the path of the charged 
particle~(see, e.g., Ref.~\onlinecite{batelaan09}). This technical difficulty
does not exist in our SCB analogue. The interference of the two different
charge states instead of two spatially separated paths 
is manifested in the qubit's
interaction with a static external field at a distance.
The elimination of the requirement of fast switching of the electric potential 
would enable much easier realization of the ideal force-free EAB effect.

\section{Conclusion.}
In conclusion, we have predicted the scalar AB effect without
a loop in a
Cooper pair box interacting with an external electric field at a distance.
The superposition of different charge states eliminates the requirement of
loop geometry for the interferometer. 
The phase shift is given by the charge-state dependence of
the field interaction energy and is universally represented by the geometric
potential. Our proposal provides an easy way to realize the
ideal EAB effect, as the setup does not require a pulse experiment for
fast switching of the potential, which has been the major technical 
obstacle for its observation.

 \bibliography{references}

\begin{figure}
\centering
\includegraphics[width=11cm]{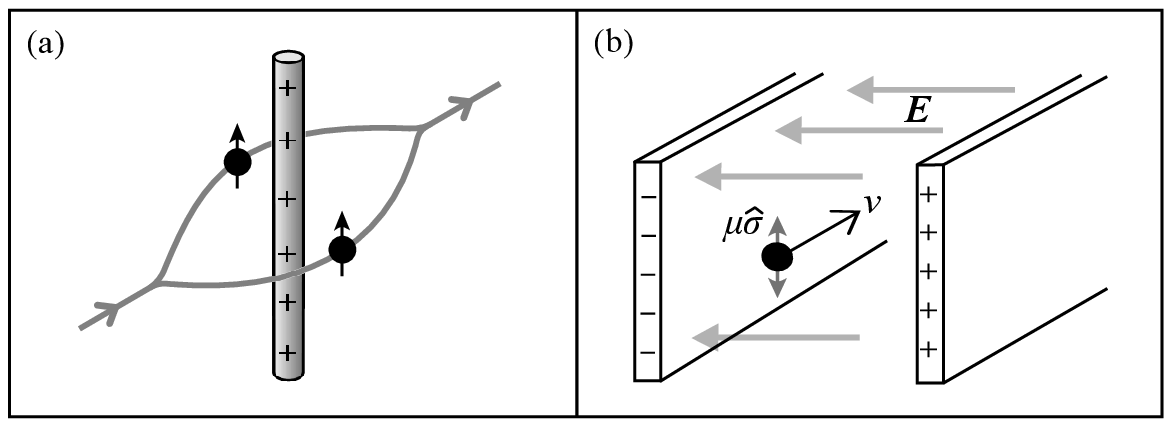}
\caption{Illustration of (a) the topological Aharonov-Casher 
 effect of a neutral particle with a fixed magnetic moment, and 
 (b) the loop-free analog of the Aharonov-Casher interference of a 
 superposed spin. 
 }
\end{figure}
%
\begin{figure}
\centering
\includegraphics[width=11cm]{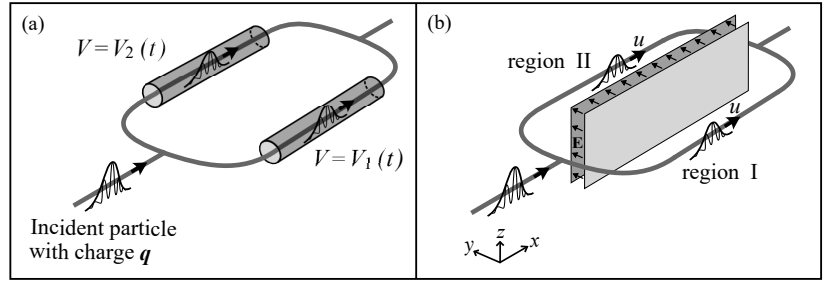}
\caption{An ideal setup of the force-free electric Aharonov-Bohm effect: 
 (a) the original
 arrangement of Aharonov and Bohm~\cite{aharonov59}
 and (b) an equivalent situation
 with the potential difference represented by the 
 electric field localized between two paths. 
 }
\end{figure}
\begin{figure}
\centering
\includegraphics[width=13cm]{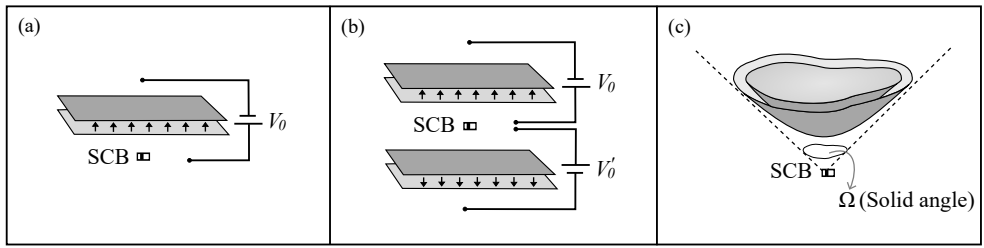}
\caption{Various configurations of the electric field at a distance from the
 location of the Cooper pair box. 
 The electric Aharonov-Bohm effect can be tested
 with these arrangements, each of which yields a different 
 geometric potential: 
 (a) $V_G = V_0/2$, (b) $V_G = (V_0+V_0')/2$, 
and (c) $V_G = V_0\Omega/4\pi$,
 where $\Omega$ is the solid angle formed in the configuration of the 
 external field.
 }
\end{figure}
%
\begin{figure}
\centering
\includegraphics[width=8cm]{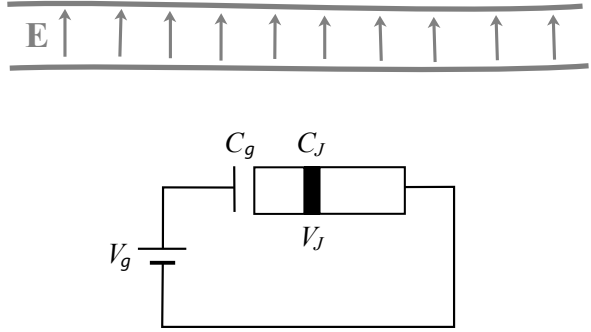}
\caption{A single Cooper pair box (SCB) circuit 
in the presence of an arbitrary 
distribution of the external electric field spatially separated 
from the circuit. 
The setup is equivalent to the standard SCB circuit, except that
the distant electric field distribution gives rise to an additional 
field interaction energy.
 }
\end{figure}
\begin{figure}
\centering
\includegraphics[width=8cm]{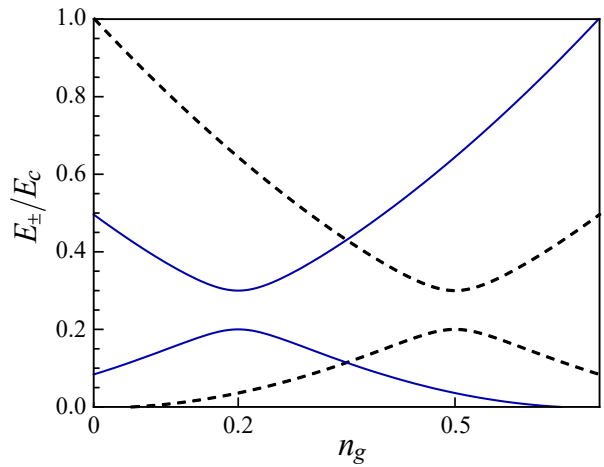}
\caption{Electric Aharonov-Bohm effect in the spectrum of an SCB.
 Solid (Dashed) lines represent the gate dependence of
 the qubit eigenstate energies
 for $2eV_G/E_C = 0.6$ ($2eV_G = 0$) and $E_J/E_c=0.1$. 
 The shift of the spectrum at nonzero $V_G$ results from the interaction
 of the SCB charge with the external electric field at a distance.
 }
\end{figure}
\begin{figure}
\centering
\includegraphics[width=8cm]{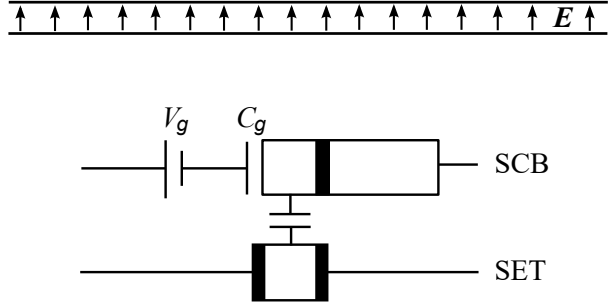}
\caption{A possible measurement scheme of the qubit state with a single 
 electron transistor capacitively coupled to the SCB.
 This measurement can probe the electric Aharonov-Bohm effect manifested
 in the qubit state.
 }
\end{figure}
\begin{figure}
\centering
\includegraphics[width=10cm]{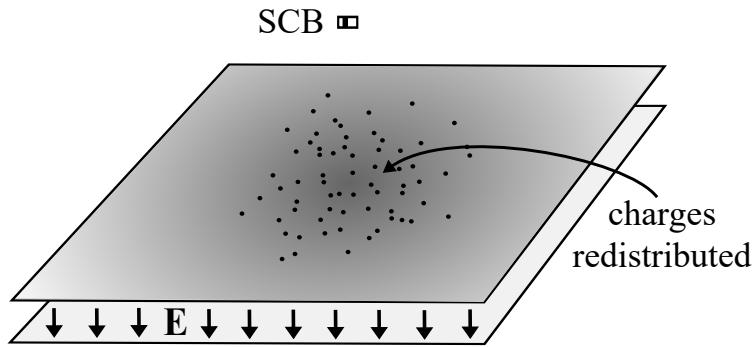}
\caption{Illustration of the charge redistribution 
on the surface of the conducting plate induced by the SCB charge $q$.
 }
\end{figure}

\end{document}